\documentclass[a4paper]{jpconf}
\usepackage[latin1]{inputenc} \usepackage{amsmath}
\usepackage{color} \usepackage{hyperref} \usepackage{xspace}
\usepackage{amsfonts} \usepackage{amssymb} \usepackage{graphicx}

\newcommand{\dd}{{\rm d}} 

\newcommand{\UNIT}[1]{\ensuremath{\,{\rm #1}}\xspace}

 \newcommand{\GeV}{\UNIT{GeV}}
 
 \newcommand{\TeV}{\UNIT{TeV}}
 
 \newcommand{\fm}{\UNIT{fm}}
 
\newcommand{\fmc}{\ensuremath{\,{\rm fm}/c}\xspace}

\definecolor{magenta}{cmyk}{0,1,0,0}

\begin{document}
% \makeatletter

\title{Remarkable Features of Decaying Hagedorn States}

\author{M Beitel, K Gallmeister and C Greiner}
\address{Institut f\"ur Theoretische Physik, Goethe-Universit\"at
  Frankfurt am Main, Max-von-Laue-Str.~1, 60438 Frankfurt am Main,
  Germany}
\ead{beitel@th.physik.uni-frankfurt.de}

\begin{abstract}
Hagedorn states (HS) are a tool to model the hadronization process which occurs 
in the phase transition phase between the quark gluon plasma (QGP) and the 
hadron resonance gas (HRG). Their abundance is believed to appear near the 
Hagedorn temperature $T_H$ which in our understanding equals the critical 
temperature $T_c$. These hadron-like resonances are characterized 
by being very massive and by not being limited to quantum numbers of known 
hadrons. To generate a whole zoo of such new states we solve the covariantly 
formulated bootstrap equation by regarding energy conservation and conservation 
of the baryon number $B$, strangeness $S$ and electric charge $Q$. To investigate 
their decay properties decay chain calculations of HS were conducted. 
One single (heavy) HS with certain quantum numbers decays by various 
two-body decay channels subsequently into final stable hadrons. Multiplicities of 
these stable hadrons, their ratios and their energy distributions are presented. 
Strikingly the final energy spectra of resulting hadrons show a thermal-like 
distribution with the characteristic Hagedorn temperature $T_H$. All hadronic 
properties like masses, spectral functions etc. are taken from the hadronic transport 
model Ultra Relativistic Quantum Molecular Dynamics (UrQMD).
\end{abstract}

\section{Introduciton}
Before the emergence of quantum chromodynamics (QCD) as the theory of strong interactions 
many phenomenological ideas came up trying to describe particle production in elementary but 
also in heavy ion collisions. The most important idea for this note goes back to R.~Hagedorn 
\cite{Hagedorn:1965st} who proposed in 1965 that all particles found at that time and which 
would be found in the future belong to a common mass spectrum. This spectrum, better known 
as Hagedorn spectrum, exhibits the specific feature of being exponential in the infinite mass 
limit. The slope of Hagedorn spectrum's exponential part is solely determined by the 
Hagedorn temperature $T_H$. This temperature denotes the limiting temperature for hadronic 
matter since any partition function of a HRG with Hagedorn-like mass spectrum diverges as long 
as $T>T_H$. Above the Hagedorn temperature a new state of matter, namely the QGP, is assumed 
to be realized. How a phase transition from HRG to QGP and back exactly works is one of the 
most challenging problems of modern physics. One possible tool to investigate this phase transition 
is the application of HS which mainly contribute to the exponential part of the Hagedorn
spectrum but also may appear in the 'hadronic' mass range too. The HS are created 
in multi-particle collisions most abundantly near $T_H$ which in our understanding equals to the critical 
temperature $T_c$. Hagedorn states are color neutral objects which are allowed to have any quantum 
numbers as long as they are compatible to HS' mass. The appearance of HS 
in multi-particle collisions and their role was already discussed in 
\cite{Greiner:2000tu,Greiner:2004vm,NoronhaHostler:2007jf,NoronhaHostler:2010dc,
NoronhaHostler:2009cf}. In 
\cite{Moretto:2005iz,Zakout:2006zj,Zakout:2007nb,Ferroni:2008ej,Bugaev:2008iu,Ivanytskyi:2012yx}.
the authors show that HS can also alter the occurrence of 
various phases from hadronic to deconfined partonic matter (first order, second order or crossover) improving  QCD's  
equation of state as shown in \cite{NoronhaHostler:2008ju,Majumder:2010ik,Karsch:2003vd}. 
The appearance of HS near $T_c$ can explain,
as shown in \cite{NoronhaHostler:2007jf,NoronhaHostler:2010dc,NoronhaHostler:2009cf},
the fast chemical equilibration of (multi-) strange baryons $B$ and their anti-particles $\bar{B}$ at 
Relativistic Heavy Ion Collider (RHIC) energies considering HS as a kind of catalyst according to
\begin{equation}\label{eq:rateq}
	\left(n_1\pi+n_2K+n_3\overline{K}\leftrightarrow\right)HS\leftrightarrow\overline{B}+B+X.
\end{equation}
The dynamical evolution of this reaction is given through a set of coupled rate equations leading to 
a chemical equilibration time of about $t_{ch}\approx5\fmc$ for $B\bar{B}$-pairs. Without 'clustering' 
of pions and kaons to HS the same approach would result at least in $t_{ch}\approx10\fmc$ or more which is obviously too 
long. The inclusion of HS in a hadron resonance gas model provides
also a lowering of the speed of sound, $c_s$ and of the shear viscosity over entropy density ratio 
$\eta/s$ at the phase transition and
being in good agreement with lattice calculations
\cite{NoronhaHostler:2008ju,Majumder:2010ik,NoronhaHostler:2012ug,Jakovac:2013iua}.
In addition, by comparing calculations with inclusion of HS to calculations without them, a significant lowering of the
shear viscosity to entropy ratio, $\eta/s$, is observed
\cite{NoronhaHostler:2008ju,Gorenstein:2007mw,Itakura:2008qv,NoronhaHostler:2012ug}.
The inclusion of HS creates a minor dependence of the
thermal fit parameters of particle ratios on the Hagedorn temperature,
$T_H$, which is assumed to be equal to $T_C$
\cite{NoronhaHostler:2009tz}. The successful application of HS mentioned above calls 
to an implementation of them into the hadronic transport program UrQMD on the basis of 
$2\leftrightarrow 1$-processes by regarding the principle of detailed balance. The role of HS on 
dynamical evolution of multiplicities, transport coefficients etc.~will thus be investigated in future \cite{Beitel2014}.
\section{Model}
The formulation of a whole zoo of HS will be provided as they will be created in binary collisions within the
microscopic hadronic transport simulation program UrQMD
\cite{Bass:1998ca}. Multiplicities (and their ratios) of stable
hadrons stemming from cascading decay simulations of one single initial massive 
HS for different masses, radii and quantum number combinations  
are calculated. Additionally energy distribution of the decay
products are examined and it is shown that all hadrons stemming from
that cascade follow the Boltzmann distribution analogous alike a
thermalized hadron resonance gas, although the final and freely moving hadrons are freed solely from the subsequent decay. 
The starting point of all calculations provided is the postulate of the statistical bootstrap model (SBM) 
stating that fireballs consist of fireballs which in turn consist of fireballs etc.~. The mathematical
formulation of this postulate leads to the well known bootstrap equation
\begin{align}\label{eq:tautotbsq}
  \tau_{\vec{C}}\left(m\right)&=\frac{R^3}{3\pi
m}\sum\limits_{\vec{C}_1,\vec{C}_2}\iint
	\dd m_1\dd m_2\,\tau_{\vec{C}_1}(m_1)\,m_1\tau_{\vec{C}_2}(m_2)\,m_2\\
  &\times p_{cm}\left(m,m_1,m_2\right)\delta_{\vec{C},\vec{C}_1+\vec{C}_2}\
  \nonumber
\end{align}
The functions $\tau_{\vec{C}_i}$ on the r.h.s.~are spectral functions of two constituents which make up 
a HS with spectral function $\tau$ on the l.h.s.~ of Eq.~(\ref{eq:tautotbsq}). Strict conservation 
of total energy leads $p_{cm}$ denoting the momenta of both constituent particles with masses $m_1$ and $m_2$ in 
the rest frame of made up HS with mass $m$,
\begin{align}\label{eq:pcm}
	p_{cm}\left(m,m_1,m_2\right)=\frac{1}{2m}\sqrt{\left(m^2-m_1^2-m_2^2\right)^2-4m_1^2m_2^2},
\end{align}
as usual where charge conservation is assured by Kronecker's $\delta$. The radius $R$ denotes the size of created 
HS and is considered to be constant taking on some reasonable values which are discussed further below. 
Contrary to the well-known non-covariant bootstrap equation \cite{Frautschi:1971ij,Hamer:1971zj}, 
the expression here is formulated covariantly. In the general solution of Eq.~(\ref{eq:tautotbsq}), 
the number of constituents is theoretically infinite. The reason to consider two constituents 
case only is because HS will be implemented in hadron transport models like e.g.~UrQMD as a whole zoo of new particles. 
In standard transport models maximally two particles in the incoming channel are allowed because the 
interaction probability is calculated on the basis of geometrical cross sections. On the other hand resonance decays 
in two hadrons are realized in UrQMD too making an implementation of further 
$\left(2\leftrightarrow 1\right)$ processes, now involving HS, possible. For this kind 
of new processes the principle of detailed balance will strictly hold. The accepted 
error by the approximation of only two outgoing particles is roughly about $30\%$, which can be estimated by looking at the HS decay probability into $n$ particles,
$P\left(n\right)=\left(\ln 2\right)^{n-1}/\left(n-1\right)!$, yielding
a probability for the decay into two particles of $69\%$, into three
particles of $24\%$ etc.~\cite{Frautschi:1971ij}. 
Given the function $\tau\left(m\right)$ we proceed to the formulation of HS' total decay width,
\begin{align}\label{eq:gamgen}
	\Gamma_{\vec{C}}\left(m\right)&=\frac{\sigma}{2\pi^2\tau_{\vec{C}}\left(m\right)}\sum\limits_{\vec{C}_1,\vec{C}_2}\iint\dd m_1\dd m_2\tau_{\vec{C}_1}\left(m_1\right)\tau_{\vec{C}_2}\left(m_2\right)\nonumber\\
  &\times
  p_{cm}^2\left(m,m_1,m_2\right)\delta_{\vec{C},\vec{C}_1+\vec{C}_2}\
  .
\end{align}
This formula was derived by employing general formulae for cross section and decay width as given in \cite{PhysRevD.86.010001ab} where for 
the (creation or fusion) cross section $\sigma$ of HS only the $2\rightarrow1$ and for their decay width $\Gamma$ only the $1\rightarrow2$ case were 
considered. Further we demanded the principle of detailed balance between creation and decay of HS to be valid which connects its creation with 
its decay properties in Eq.~(\ref{eq:gamgen}). The cross section will be considered simply as the geometrical value $\sigma=\pi R^2$. The partial decay width of a HS with mass $m$ and charge 
vector $\vec{C}$ decaying into two particles with masses between $m_1$ and $\dd m_1+m_1$, $m_2$ and $\dd m_2+m_2$  and charges $\vec{C}_1$, $\vec{C}_2$ reads
\begin{align}\label{eq:gamgenpar}
	\Delta\Gamma_{\vec{C},\vec{C}_1,\vec{C}_2}\left(m,m_1,m_2\right)&=\frac{\sigma}{2\pi^2}\frac{\Delta m_1\tau_{\vec{C}_1}\left(m_1\right)\Delta m_2\tau_{\vec{C}_2}\left(m_2\right)}{\tau_{\vec{C}}\left(m\right)}p_{cm}^2\left(m,m_1,m_2\right)
\end{align}
The fractional two-body branching ratios $\mathcal{B}$ are just the ratio of partial and total decay widths, Eqs.~(\ref{eq:gamgenpar}) and (\ref{eq:gamgen}),
\begin{align}\label{eq:bra}
	\Delta\mathcal{B}_{\vec{C},\vec{C}_1,\vec{C}_2}\left(m,m_1,m_2\right)=\frac{\Delta\Gamma_{\vec{C},\vec{C}_1,\vec{C}_2}\left(m,m_1,m_2\right)}{\Gamma_{\vec{C}}\left(m\right)}\ ,
\end{align}

%%% RESULTS %%%
\section{Results}
\begin{figure}
\begin{minipage}[c]{.8\textwidth}
\includegraphics[scale=0.25,angle=270]{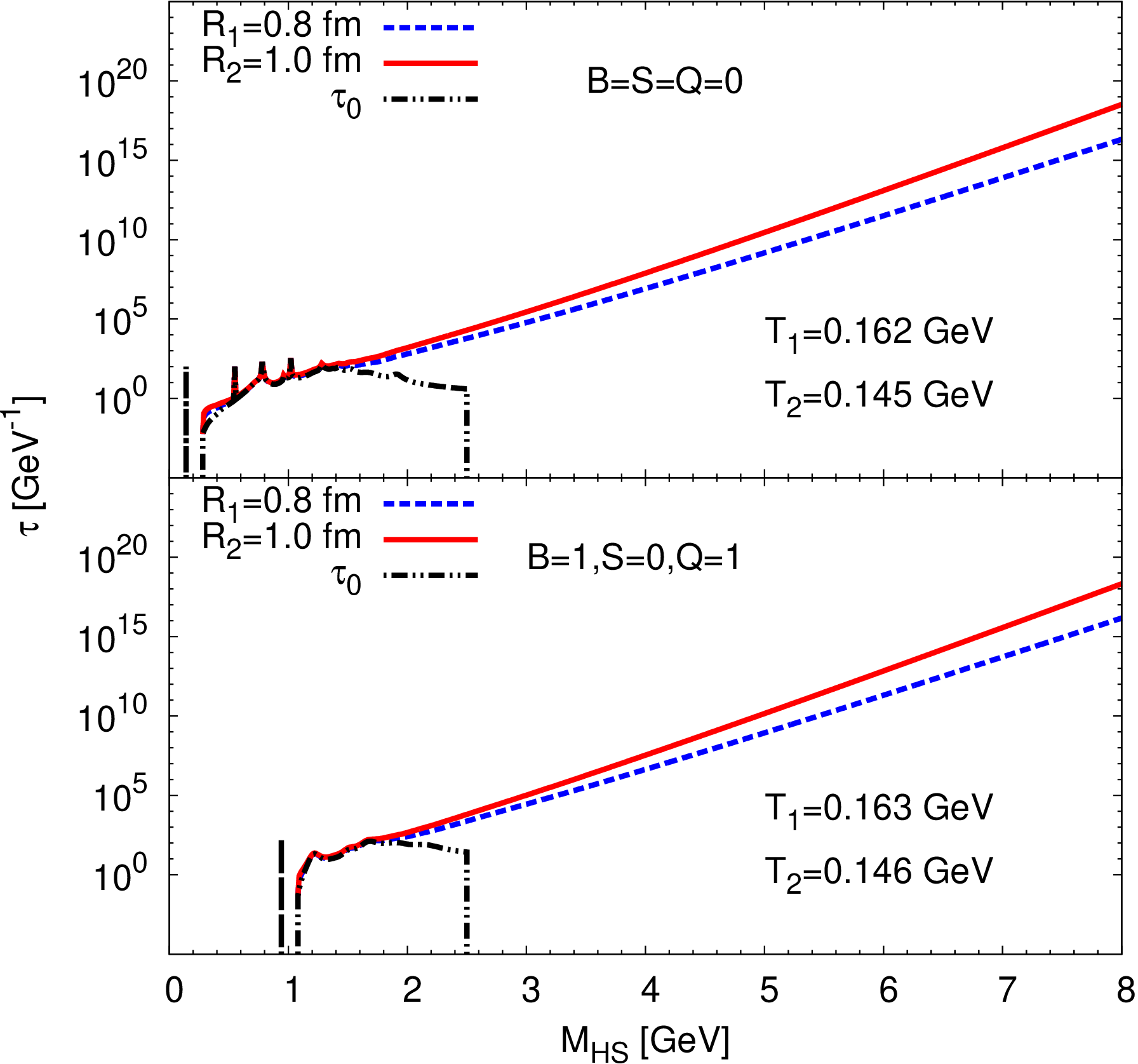}
\includegraphics[scale=0.25,angle=270]{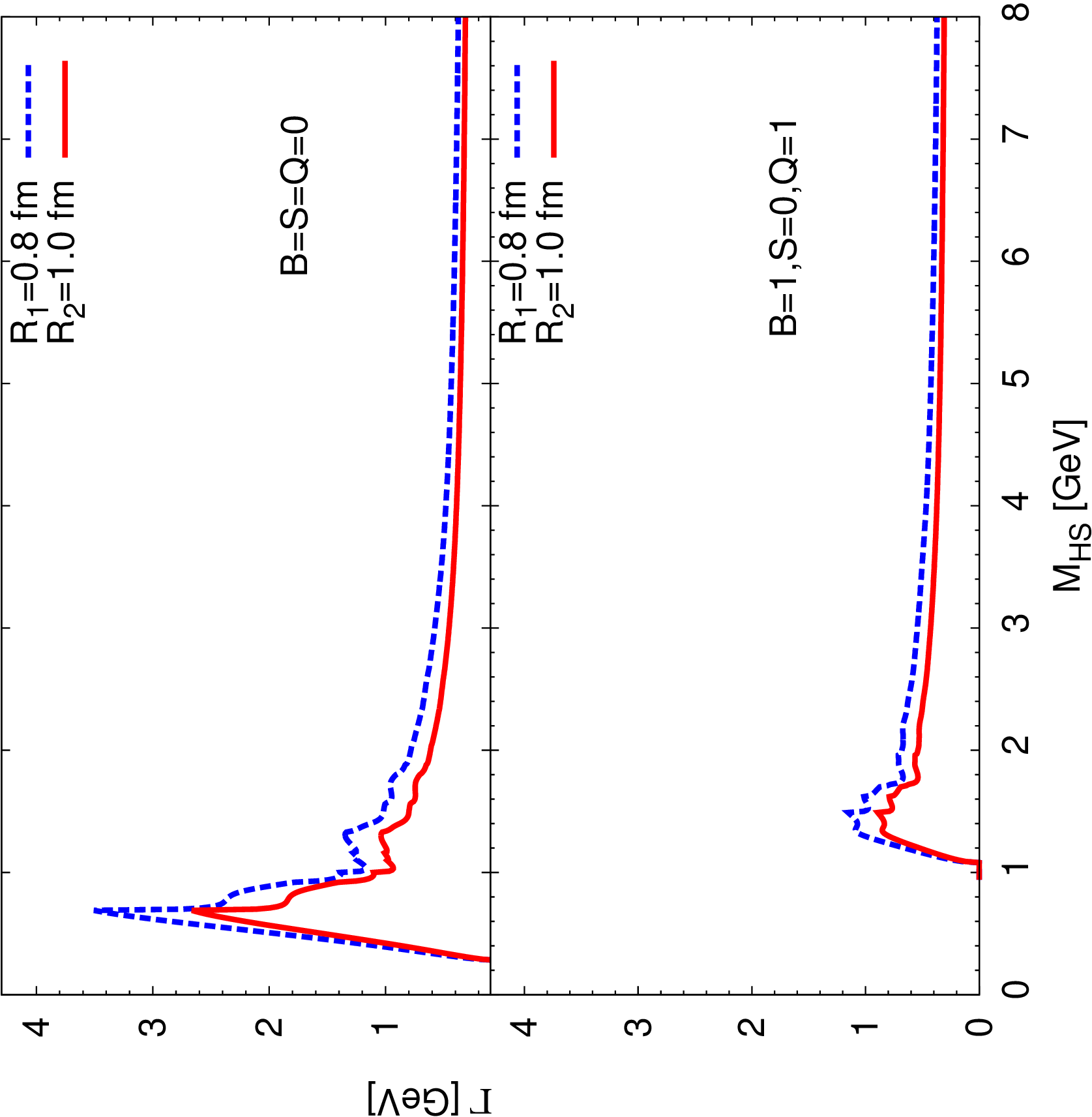}
\end{minipage}
  \caption{Meson-like $\left(B=S=Q=0\right)$ Hagedorn spectrum (upper left) and 
	corresponding HS' total decay width (upper right) and baryonic
    $\left(B=1,S=0,Q=1\right)$ Hagedorn spectrum (lower left) and 
	corresponding HS' total decay width (lower right) for two
    different radii. On the left figure additionally (fitted) Hagedorn
    temperatures are provided where the black line represents the sum of spectral
    functions of hadrons with the given quantum numbers.}
  \label{fig:taugam}
\end{figure}

\begin{figure}
  \begin{minipage}{.8\textwidth}
  \includegraphics[width=0.45\textwidth,angle=270]{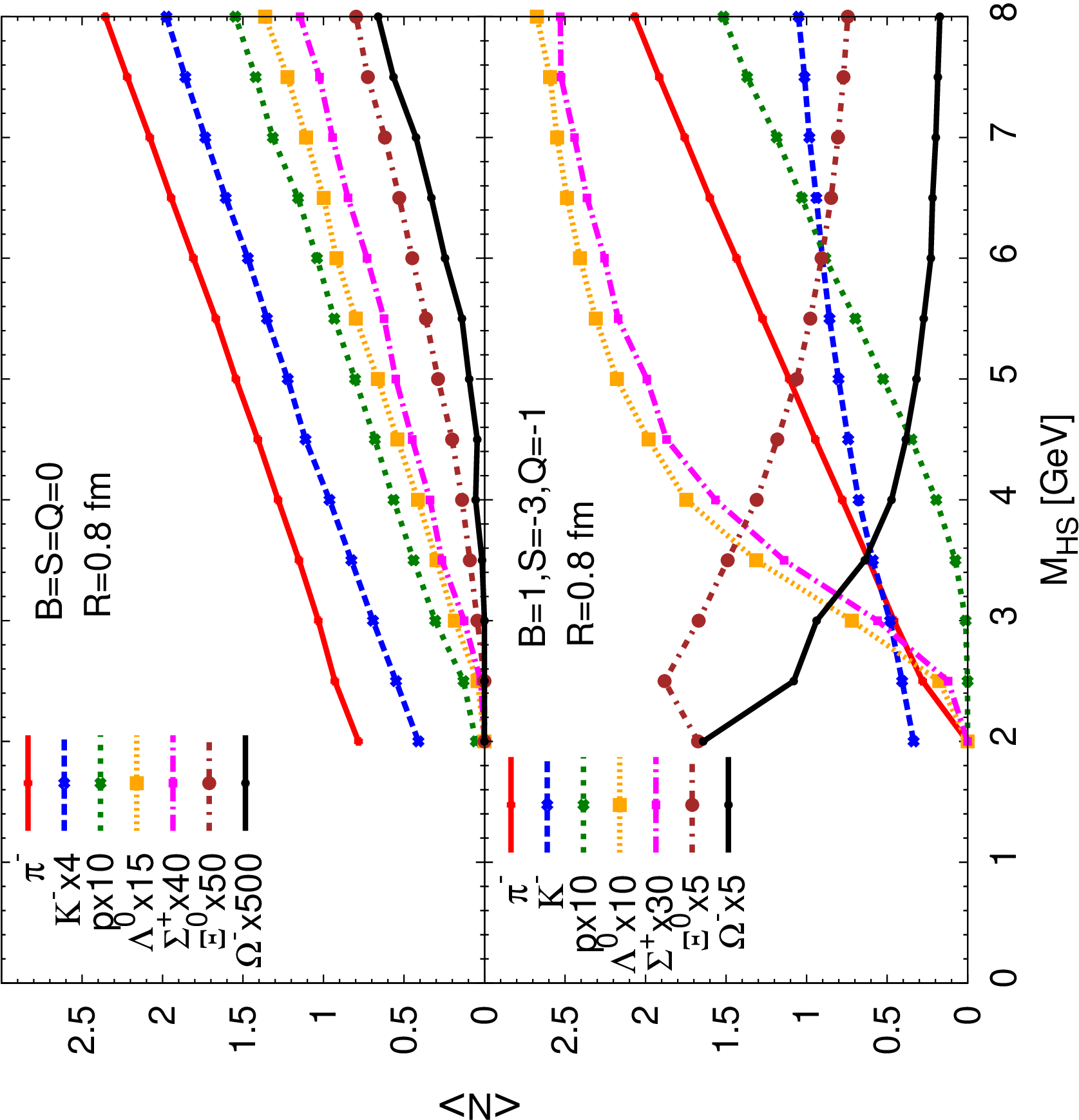}
  \includegraphics[width=0.45\textwidth,angle=270]{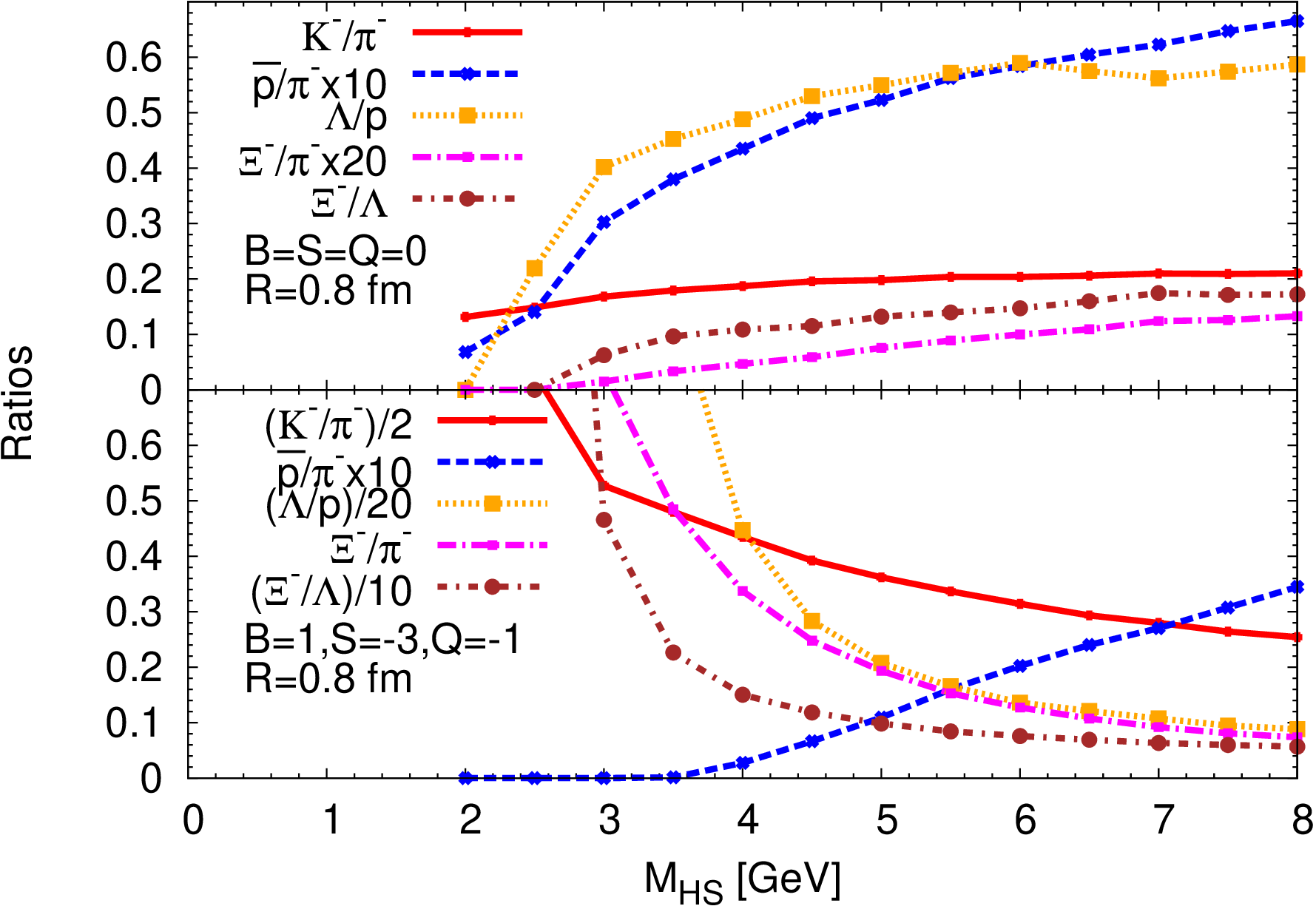}
\end{minipage}
  \caption{Hadronic multiplicities after a cascade decay of initial Hagedorn
    state with radius $R=0.8\fm$ and $B=S=Q=0$ (upper left) and $B=1,S=-3,Q=-1$ (lower left). 
    Corresponding multiplicity ratios are shown for $B=S=Q=0$ (upper right) and $B=1,S=-3,Q=-1$ 
    (lower right). In both cases 'hadronic' feeddown is taken into account.}
  \label{fig:mulprlst}
\end{figure}

\begin{figure}
  \centering
  \includegraphics[scale=0.30,angle=270]{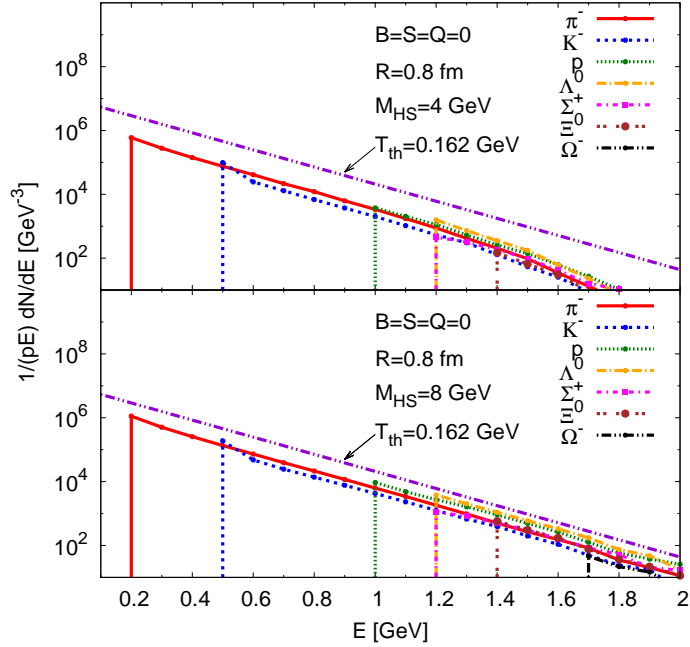}
  \caption{Energy spectra of hadrons stemming from cascade decay of
    charge neutral Hagedorn state with radius $R=0.8\fm$ and initial
    mass $M_{HS}=4\GeV$ and $M_{HS}=8\GeV$.}
  % \NOTE{1.0\fm}
  \label{fig:eneprl}
\end{figure}

The bootstrap equation Eq.~(\ref{eq:tautotbsq}) in general is a highly
non-linear integral equation of Volterra type which can be solved
analytically for some special cases
\cite{Yellin:1973nj,Hagedorn:1973kf}. The numerical solution of the given bootstrap equation for a meson-like,
non-strange and electrically neutral (B=S=Q=0) Hagedorn spectrum for
two different typical radii ($R_1=0.8\fm, R_2=1.0\fm$) is presented in
Fig.~\ref{fig:taugam}.  In the same figure also spectra for baryonic
non-strange and electrically charged states $\left(B=1,S=0,Q=1\right)$
are shown.
All Hagedorn spectra rise exponentially for masses $\geq 1.5\GeV$ with
different slopes for different radii, but for $m<1.5\GeV$ they all
include and thus fit the 'hadronic' part of the spectrum. The slopes of the exponential part were determined by the fit function $\tau_{\rm
  fit}\left(m\right)=Am^{-b}\exp\left(m/T_H\right)$, yielding the Hagedorn temperatures
$T_H=0.145\GeV$ for $R=1.0\fm$ and $T_H=0.162\GeV$ for $R=0.8\fm$ both being rather independent on the chosen
quantum number combinations.
\begin{figure}
\centering
\includegraphics[width=0.8\textwidth,angle=0]{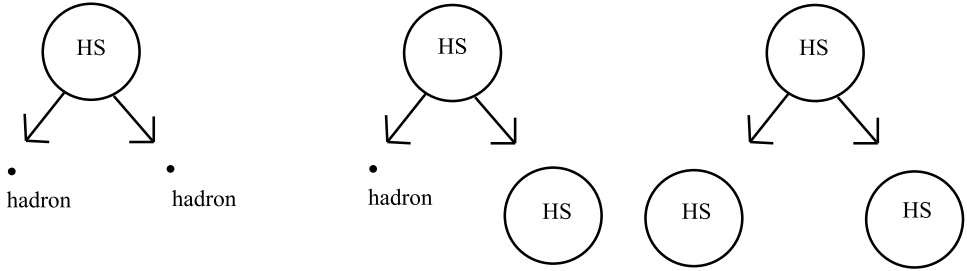}
\caption{Decay of HS into two hadrons (left), decay into one hadron and one HS (center) and decay into two HS (right)}
\label{fig:deccart}
\end{figure}
\begin{figure}
\centering
\includegraphics[width=0.5\textwidth,angle=0]{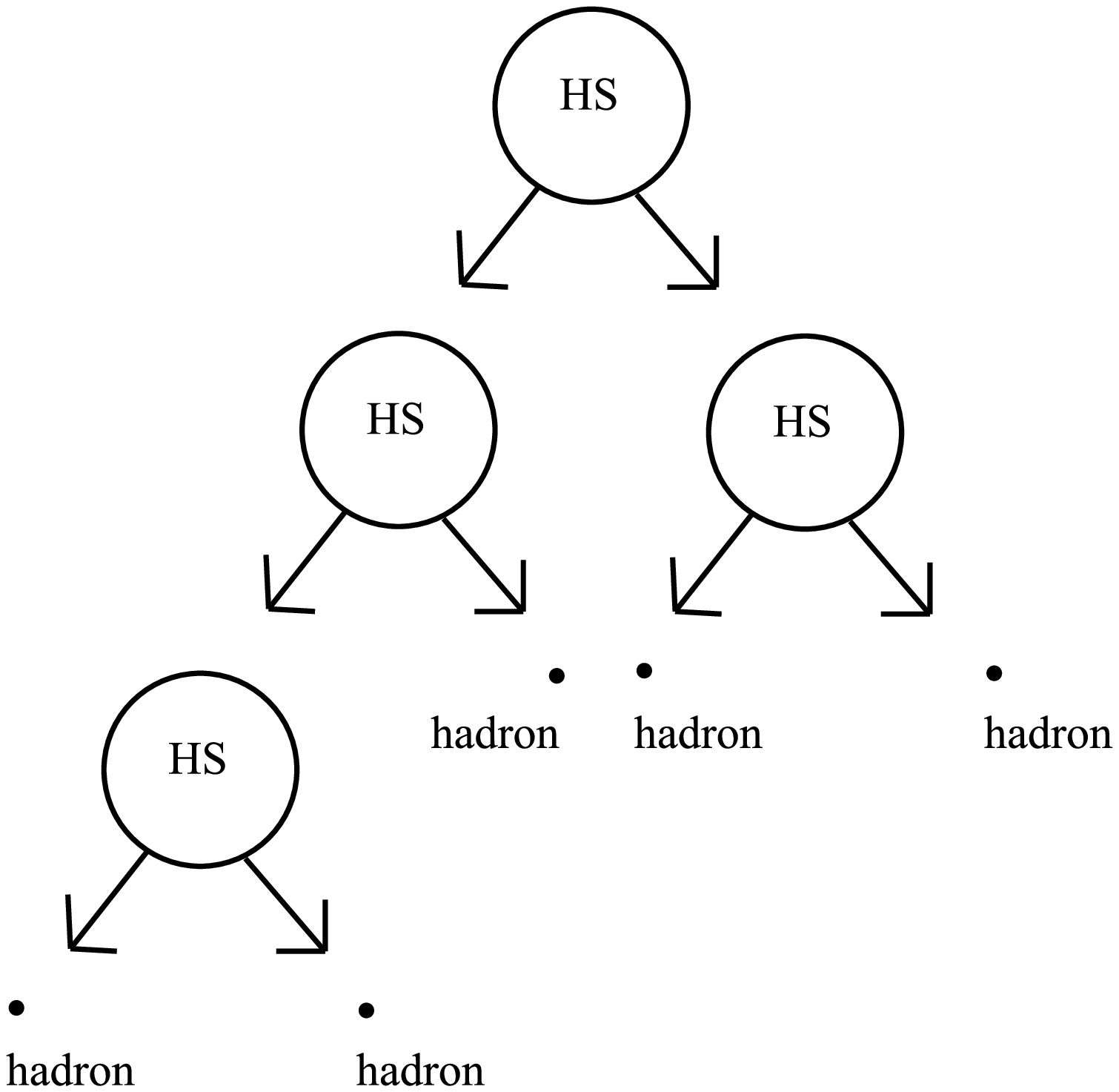}
\caption{One possible decay chain of an initial heavy HS}
\label{fig:cascart}
\end{figure}
The total decay width of a Hagedorn state consists of three different
contributions, where the first one considers only hadrons, 
the second hadrons and Hagedorn
states, and the third one only
HS in the outgoing
channel as depicted in Fig.~\ref{fig:deccart}. The peak in the mass range of $M_{HS}=0$-$2\GeV$ 
on left part of Fig.~\ref{fig:taugam} comes mainly
from the first contribution, because in this mass range the phase
space for pure hadronic decay is largest.
The height of the peak
depends on the number of hadronic pairs, whose quantum numbers all sum
up to the quantum number of the Hagedorn state they are building up,
being large for $B=S=Q=0$ and rather small for $B=Q=1,S=0$. Another remarkable feature is that for both
radii the total decay width tends to a constant value depending only on $R$ for large masses. 
Having the numerous branching ratios Eq.~(\ref{eq:bra}) at hand, one is able to
calculate hadronic multiplicities stemming from Hagedorn state decays.
Here one starts with some initial heavy Hagedorn state,
which decays subsequently down until hadrons are left only as shown in Fig.~\ref{fig:cascart}. Among
those also non stable resonances might appear, which further undergo a
hadronic feed down leaving one with light and stable hadrons with
respect to the strong force like pions, kaons, etc.~. All hadronic
properties used here were taken from the transport model UrQMD
\cite{Bass:1998ca}. Calculated multiplicities and their ratios for some uncharged
$\left(B=S=Q=0\right)$ initial Hagedorn state are shown in 
Fig.~\ref{fig:mulprlst} (upper part).
One observes a linear dependence of all multiplicities on the initial
Hagedorn state mass where the magnitude depends on the available phase
space for each hadron. Thus in a decay of a charge neutral Hagedorn
state $\pi^-$ clearly expectedly dominate which have to be produced in pairs mostly with
$\pi^+$ since exact charge conservation is enforced. Kaons, especially
$K^-$, are even stronger suppressed not only of their larger mass but
also due to the fact that they have to conserve both electric charge
and strangeness. For the baryons presented the same argumentation
holds since both have to conserve baryon number $B$ and additionally
electric charge $Q$ for proton and strangeness $S$ for $\Lambda$. For
the multistrange hyperons $\Xi^0$ and $\Omega^-$ the production
suppression is even stronger. This has to be contrasted with the
results for a baryonic, multi-strange and electrically charged
$\left(B=1,S=-3,Q=-1\right)$ $\Omega^-$-like Hagedorn state also shown
in Fig.~\ref{fig:mulprlst}. Now the choice of Hagedorn state's initial quantum numbers is
reflected in the preference of baryon production although they are
much heavier than the presented mesons. Especially the abundance of
hyperons $\left(\Omega^-,\Xi^0\right)$ compared to the case discussed
before is striking since the easiest way to conserve the initial
quantum numbers is the production of one $\Omega^-\pi^0$- or one
$\Xi^0K^-$ pair where on the other hand the phase space for all other
hadrons with different quantum numbers is suppressed now. Hence exact
conservation of quantum numbers always causes a competition between
hadron's phase space and its quantum numbers. On the right part of Fig.~\ref{fig:mulprlst} the 
corresponding multiplicity ratios for $R=0.8\fm$ and same quantum number combinations are shown. 
A comparison of theoretical ratios with experimental results is also provided. 
Numerical values for the multiplicity ratios for Hagedorn state masses
of 4\GeV and 8\GeV are listed in Tab.~\ref{tab:ratios} and, for illustration, compared
to experimental results for p-p- and Pb-Pb collisions at midrapidity, both measured by ALICE at LHC. 
\begin{table}[h]
	\begin{center}	
\begin{tabular}{lcccc} 
\br 
&p-p&Pb-Pb & $4\GeV$ & $8\GeV$\\ 
\mr
$K^-/\pi^-$ & 0.123(14)&0.149(16) & 0.187 & 0.210 \\ 
$\overline p/\pi^-$ & 0.053(6)&0.045(5) & 0.043 & 0.066 \\ 
	$\Lambda/\pi^-$ & 0.032(4)&0.036(5) & 0.021 & 0.038 \\ 
$\Lambda/\overline p$ & 0.608(88)& 0.78(12) & 0.494 & 0.579 \\ 
		$\Xi^-/\pi^-$ & 0.003(1)&0.0050(6) & 0.0023 & 0.0066  \\
$\Omega^-/\pi^-\,\cdot 10^{-3}$& ---& 0.87(17) & 0.086&0.560 \\
\br
\end{tabular} 
\caption{Comparison
  of particle multiplicity ratios from theory vs.~p-p at $\sqrt{s_{NN}}=0.9$\TeV \cite{Aamodt:2011zza} 
  and Pb-Pb at $\sqrt{s_{NN}}=2.76$\TeV \cite{Abelev:2013vea,Abelev:2013xaa,Abelev:2013zaa}, both from 
  ALICE at LHC. Calculated values are listed for Hagedorn state masses of 4\GeV and 8\GeV. Numbers in brackets
  denote the error in the last digits of the multiplicity ratios.}
\label{tab:ratios} 
\end{center}
\end{table}
The theoretical multiplicity ratios lie seemingly close to the measured by ALICE, except for the very rare multi-strange baryon $\Omega^-$. 
However, it has to be made clear, 
that the decay of HS alone is never assumed to describe the experimental data. The theoretical 
multiplicity ratios serve as a proof of reliability and reasonability of Hagedorn state's branching ratios defined in Eq.~(\ref{eq:bra}). 
This test is necessary before implementing them into a transport model. Beside multiplicities also the energy distribution of 
the decay products was examined. They are shown in Fig.~\ref{fig:eneprl} for an uncharged $\left(B=S=Q=0\right)$ 
Hagedorn state with initial mass $M_{HS}=4\GeV$ and also $M_{HS}=8\GeV$. The striking observations are that all particle 
species follow a Boltzmann-like distribution with the same slope independent on the initial Hagedorn state mass. Thus pions, kaons, protons spectra look alike stemming of a system with a temperature being $T_{th}=0.162\GeV$. We remark that the final and freely moving hadrons are freed from the subsequent decays without any reintaractions. The particular finding is that the 'thermal' temperature $T_{th}$ exactly equals the 
Hagedorn temperature $T_H$ obtained from a fit in the left part of Fig.~\ref{fig:taugam} for the case $R=0.8\fm$.
The Hagedorn temperature $T_H$ was nothing but a slope parameter to fit the exponential part of the
Hagedorn spectrum, where on the other hand $T_{th}$ is the slope ('temperature') of the finally created hadrons. 
Starting with a bootstrap formula with no introduction of any notion of temperatures at all resulted in a 'thermalized' decay 
with a slope being the Hagedorn temperature. 
\section{Conclusion}
A covariantly formulated bootstrap equation is presented which ensures energy and quantum number conservation. The solution of this 
bootstrap equation provides the Hagedorn spectra describing the 'hadronic' part adequately and being exponential for large masses. Given the 
Hagedorn spectra the total decay width formula for HS was presented which was obtained on the principle of detailed balance between their creation and their decay. The main characteristics of HS' total decay width are their large peaks for masses in the 'hadronic' range and roughly constant values for masses beyond that range. With partial and total decay width we were able to define HS' branching ratios needed for decay simulations. 
As a specific case we considered decay chains of one single initial Hagedorn state cascading down by various two (intermediate) particle decay channels until stable hadrons 
were left only. The multiplicity ratios of those stable hadrons where compared to Pb-Pb ALICE data at LHC to check the reasonability and reliability 
of HS' theoretical branching ratios. Also the energy distribution of those hadrons were examined for two initial HS' masses. The striking findings 
were that all hadrons stemming from such a decay chain, without any reinteractions, are first all thermal and second exhibit the same temperature. 
The particular feature of this 'thermal' temperature is that it equals the Hagedorn temperature gained from a fit of Hagedorn spectrum's 
exponential part. Summarizing, such a finding gives fresh insight into the microscopic and thermal-like
hadronization in ultrarelativistic e$^+$-e$^-$- (see
eg.~\cite{Becattini:1995if}), hadron-hadron-, and also especially in
heavy ion collisions: An implementation of the presented Hagedorn state
decays in addition to their production mechanisms into the transport
approach UrQMD offers a new venue for allowing strongly interacting hadronic multiparticle
collisions in a consistent scheme being important in the vicinity of
the deconfinement transition by creating and decaying more exotic HS. Understanding faster thermalization and
chemical equilibration, but also microscopic transport properties can
be thoroughly investigated in future \cite{Beitel2014}.
\section{Acknowledgements}
The authors acknowledges discussions with K.~Bugaev. This work was supported by the Bundesministerium f\"ur Bildung und
Forschung (BMBF), the HGS-HIRe and the Helmholtz International Center
for FAIR within the framework of the LOEWE program launched by the
State of Hesse. Numerical computations have been performed at the
Center for Scientific Computing (CSC).

\section*{References}
\bibliographystyle{iopart-num}
\bibliography{reference}

\end{document}